\documentclass[aps, nofootinbib, prd, superscriptaddress,eqsecnum,showkeys,preprintnumbers,altaffilletter]{revtex4-1}

\usepackage{amsfonts,amsmath,amssymb,amsthm,bbm,hyperref}
\usepackage{mathbbol}
\usepackage[T1]{fontenc}
\usepackage[utf8]{inputenc}
\usepackage{mathtools}
\usepackage{graphicx}% Include figure files
\usepackage{dcolumn}% Align table columns on decimal point
\usepackage{bm}
\usepackage{xcolor}
\usepackage[normalem]{ulem}
\usepackage{mathrsfs}
\usepackage{verbatim}

\newcommand*{\id}{{\normalfont\hbox{1\kern-0.15em\vrule width .8pt depth-.5pt}}}

\usepackage{graphicx}% Include figure files

\begin{document}

\title[Sample title]{Scalar perturbations of Galileon cosmologies in the mechanical approach within the late Universe}% Force line breaks with \\

\author{Jan Nov\'ak}
 \affiliation{jan.novak@johnynewman.com, Physics Department, Technical University of Liberec, Studentsk\'a 1402/2, 461 17, Liberec, Czech republic}%Lines break automatically or can be forced with \\

\begin{abstract}
We investigate the Universe at the late stage of its evolution and inside the cell of uniformity 150-370 Mpc. We consider the Universe to be filled at these scales with dust like matter, a minimally coupled Galileon field and radiation. We use the mechanical approach. Therefore, the peculiar velocities of the inhomogeneities as well as the fluctuations of the other perfect fluids can be considered nonrelativistic. Such fluids are said to be coupled because they are concentrated around the inhomogeneities. We investigate the conditions under which the physical Galileon field, i.e. compatible with results of the latest gravitational wave experiments GW150914, GW151226, GW170104, GW170814,  GW170817 and GW170608, can become coupled. We know that at the background level coupled scalar fields behave as a two-component perfect fluids: one which mimics a network of frustrated cosmic string and an another one which corresponds to an effective cosmological constant. We found a correction for the Galileon field, which behaves like a matter component.  
\end{abstract}

\pacs{Valid PACS appear here}% PACS, the Physics and Astronomy
                             % Classification Scheme.
%Use showkeys class option if keyword
                              %display desired
\maketitle

\section{\label{sec:level1}Introduction}

The $\Lambda$CDM model is consistent with observational data but the energy scale of dark energy is too low \cite{Weinberg, Starobinsky}. Therefore, this cosmological constant is not compatible with the cosmological constant originated from the vacuum energy in a quantum field theory. Already many models have been proposed to explain the present accelerated expansion of the Universe. If the history of particle physics is any guide, then one can assume that the dark energy is due to a new field. Within this setup, the most popular ones are the models based on scalar fields, \cite{Amendola}. 

Dark energy models based on minimaly coupled scalar field are named quintessence \cite{Lopez}. In this case equation of state (EoS) parameter $w\in (-1,0)$. It is as well possible to have phantom scalar fields where $w<-1$. Finally quintom models are based on scalar fields, where there is a crossing of the value $w=-1$. In addition, a dynamical EoS parameter, $w$, can help to solve the coincidence problem.  

Another possibility to describe the late-time acceleration of the Universe is to modify the law of gravity from general relativity (GR) at large distances \cite{Faraoni, Capozziello}. There have been two main approaches. The first one is to introduce a Lagrangian for gravity constructed from the Ricci, Riemann and metric tensors. Examples are $f(R)$ theories or Gauss-Bonnet gravities. The second possibility is to build up the laws of gravity from higher dimensional models that realize the cosmic acceleration through gravitational leakage to extra dimensions. The DGP (Dvali-Gabadadze-Poratti) model belongs to this class\footnote{ Modified gravity models as a mean to explain the late-time  acceleration of the universe need to be constructed to recover GR behavior in the regions of high density for the consistency with local gravity experiments. There are two possible ways to do it: the chameleon mechanism and Vainshtein mechanism, Ref. \cite{Khoury, Babichev}; } \cite{Chimento}. 

Mostly inspired by the DGP model, the authors of Ref. \cite{Rattazzi} derived the five Lagrangians that lead to field equations invariant under the Galileon symmetry $\partial_{\mu}\phi\rightarrow \partial_\mu\phi + b_{\mu}$ (the vectorial parameter $b_{\mu}$ corresponds to a constant shift) in a Minkowski spacetime:   

\begin{gather}
L_1 = M^3\phi,\ L_2= (\nabla\phi)^2,\ L_3 = (\square\phi)(\nabla\phi)^2/M^3,\nonumber\\
L_4 = (\nabla\phi)^2\big[2(\nabla\phi)^2 - 2\phi_{;\mu\nu}\phi^{;\mu\nu} - R(\nabla\phi)^2/2\big]/M^6,\nonumber\\
L_5 = (\nabla\phi)^2\big[(\nabla\phi)^3 - 3 \square\phi\phi_{;\mu\nu}\phi^{;\mu\nu} + 2\phi_{;\mu}^{\ \nu}\phi_{;\nu}^{\ \rho}\phi_{;\rho}^{\ \mu}
-6\phi_{;\mu}\phi^{;\mu\nu}\phi^{;\rho}G_{\nu\rho}\big]/M^9, M^3 = M_{Pl} H_0^2\nonumber
\end{gather}

The scalar field $\phi$ that respects the Galileon symmetry is named the Galileon field \cite{Felice}. Each of the five terms give origin to a second order differential equation of motion, which keep the theory free from unstable spin-2 ghost degrees of freedom. If we carry the analysis in a curved spacetime, we need to change the Lagrangians to their covariant form\footnote{Some of the terms $L_i$ are already disfavoured by the results of the experiment GW170817, \cite{Miguel}, but the term which we will consider in our work is fully viable.}. 

It is of great importance to suggest a mechanism which can verify the viability of Galileon models. The theory of cosmological perturbations is a powerful tool to investigate cosmological models \cite{Mukhanov}. In fact, we will study the Universe at the late stages of its evolution and deep inside the cell of uniformity. In this setup, there are discrete inhomogeneities in this cell, because galaxies, group and clusters of galaxies were already formed and can be considered as discrete sources for the gravitational potential. It was shown in previous works that in this case the mechanical approach \cite{Eingorn, Zhuk, Alexander} is a powerful tool to study the scalar perturbations. It enables us to get the gravitational potential and to describe the motions of galaxies. 

The hydrodynamical approach is a good tool to investigate the growth of structure of the early universe. It works well in the linear approximation. However, it became inapplicable at the strongly nonlinear regime. It starts for $z$ of few dozens. However, on much bigger scales matter becomes on average homogeneous and isotropic with matter in the form of the perfect fluid. It is important to define theoretically at which scales we should define the transition from the highly inhomogeneous mechanical distribution to the smooth hydrodynamical one.

The Universe is filled with inhomogeneously distrubuted discrete structures at the scale 150-370 MPc. The mechanical approach enables us to obtain the gravitational potential for an arbitrary number of randomly distributed inhomogeneities. We can investigate the relative motion of galaxies and the formation of the Hubble flow with the expression for the gravitational potential.    
The mechanical approach works well for the $\Lambda$CDM model, where the peculiar velocities of the inhomogeneities could be considered as negligibly small when we compare them with the speed of light. Additionally, we consider scales deep inside the cell of uniformity. Consequently, we can drop the peculiar velocities at first order of approximation. Such models were generalized also to the case of cosmologies with different perfect fluids, which can play the role of dark energy and dark matter \cite{Burgazli, Brilenkov,  Akarsu, Morais, Perfect}.   

The fluctuations of these additional perfect fluids form also their own inhomogeneities. It is supposed in the mechanical approach that the velocities of displacement of such inhomogeneities are of the order of the peculiar velocities of the inhomogeneities corresponding to dust like matter. These types of inhomogeneities are coupled to each other in the sense mentioned in \cite{Kumar}. This means that for the considered models, we investigate the possible existence of such coupled fluids. They can as well play the role of dark matter as shown in \cite{Perfect}.  

In the present paper, we consider a cosmological model with a Galileon field minimally coupled to gravity. The Universe is also filled with dust like matter and radiation. We study the theory of scalar perturbations, Ref. \cite{Albarran}, for such models and obtain a condition under which the inhomogeneities of the dust like matter and the inhomogeneities of the scalar field can be coupled to each other. We show that this condition imposes rather strong restrictions on the scalar field itself. The coupled scalar field behaves at the background level as a three component perfect fluid: a cosmological constant, a term which mimics a network of frustrated cosmic strings and a further component which behaves as matter. Though the gravitational wave experiments constrain the Galileon cosmologies, here we provide a further tool based on the mechanical approach to choose the viable theories.  

The work is structered followingly: we present the action, which we consider in this article, in the first section; We study the mechanical approach in the next part III. And we present the results in the final part IV. 

\section{Summary of the Galileon models}
Given that non-gravitational coupling between the Galileon field and gravity are physically allowed, we will simply consider the following self-interaction term for the Galileon field:
\begin{equation}\label{action}
S_I = \alpha \int_M\sqrt{|g|} \square\phi\ \partial_{\mu}\phi\partial^{\mu}\phi\ d^4x,\nonumber
\end{equation}
On the previous expression, $\alpha$ is a small parameter, which measure the deviation from the model of a {minimally coupled scalar field} and has units of volume. We intend to demonstrate the typical behavior for Galileon models on this Lagrangian within the mechanical approach. First of all, we will obtain the tensor of energy momentum for this Lagrangian as: 
\begin{gather}
T_{\mu\nu}= \frac{2}{\sqrt{-g}}\frac{\delta S_{I}}{\delta g^{\mu\nu}},
\end{gather}
where $\frac{\delta S}{\delta g^{\mu\nu}}$ is the variational derivative of the action. We then get \cite{Quiros}, 

\begin{gather}
T_{\mu\nu} = \alpha\ 2 \partial_{\mu}\phi\ \partial_{\nu}\phi\ \square\phi - 2\alpha\ \phi_{;(\mu}\ (\partial_{\rho}\phi\ \partial^{\rho}\phi)_{;\nu)} +\ \alpha\ g_{\mu\nu}\ \phi_{;\lambda}\ (\partial_{\rho}\phi\ \partial^{\rho}\phi)^{;\lambda}.
\end{gather}

We would like to use the following perturbed metric in our computations according to \cite{Kumar}:

\begin{gather}
ds^2 = a^2\big[(1+2\Phi)\ d\eta^2 - (1-2\Psi)\ \gamma_{ij}dx^idx^j\big]
\end{gather}
For simplicity we assume spatially flat solutions $\gamma_{ij} = \gamma^{ij} = \id_{3\times3}$, therefore $\gamma = \gamma_{ij}\gamma^{ij} = 3$.

\section{Mechanical approach}
\begin{center}
Now we can write the tensor of energy-momentum for whole action, when we include also the minimally coupled scalar field: 
\begin{gather}
S = \int_M\sqrt{|g|}\big(\frac{1}{2}\partial_{\rho}\phi\partial^{\rho}\phi - V(\phi) + \alpha\square\phi\partial_{\mu}\phi\partial^{\mu}\phi\big)\ d^4x.\nonumber
\end{gather}

\end{center}

The characteristic term in Galileon cosmologies is the term $\square\phi$, so we need to compute $\square\phi$ up to first order on the perturbative setup. We use the notation $\phi = \phi_c +\varphi$, where $\phi_c$ is the background term and $\varphi$ the perturbed quantity (all derivatives will be with respect to the conformal time): 

\begin{gather}
\square \phi =  \frac{\phi_c''}{a^2} -\frac{a'\phi_c'}{a^3}+ \frac{a'\phi_c'}{a^3}\gamma +\big[ - \frac{2}{a^2}\phi_c''\Phi+\frac{\varphi''}{a^2}-\frac{\Delta_{ij}\varphi}{a^2} - \frac{\gamma}{a^2}\phi_c'( \Psi' + 2\Phi\frac{a'}{a}) - \nonumber\\- \frac{a'}{a}\frac{\gamma}{a^2}\varphi' - \frac{\varphi'a'}{a^3} 
+\frac{4\Phi\phi_c' a'}{a^3} - \frac{\phi_c'}{a^3}(\Phi'a + 2a' \Phi) - \square_{ij}\varphi\frac{1}{a^2}\big],
\end{gather}

where we have defined $\Delta_{ij}\varphi\equiv \varphi_{,ij}\gamma^{ij}$ and $-\varphi_{,k}\Gamma^{k}_{ij}\gamma^{ij} = \square_{ij}\varphi - \Delta_{ij}\varphi$. So, we can finally compute $T^{0}_{0}$ for $S_{I}$:

\begin{center}
\begin{gather}
\bar{T}^{0}_{0} + \delta T^0_{\ 0}  =2\alpha\ \frac{1}{a^2}(\phi_c')^2 \big(\frac{\phi_c''}{a^2}  +  \frac{\gamma}{a^2}\phi_c'\frac{a'}{a} - \frac{a'\phi_c'}{a^3} \big)- 2\alpha\frac{1}{a^4} (\phi_c')^2.(\phi_c'' -2 \phi_c'H + \phi_c' H)+\nonumber\\
+2\alpha\frac{(\phi_c')^2}{a^2}\big[-\frac{2}{a^2}\phi_c''\Phi + \frac{\varphi''}{a^2} - \frac{\Delta_{ij}\varphi}{a^2} + \frac{\gamma}{a^2}(-\phi_c'\Psi' - 2\Phi\frac{a'}{a}\phi_c' - \frac{\varphi' a'}{a} ) -\frac{\varphi'a' }{a^3}  + \frac{4\Phi\phi_c' a'}{a^3} - \frac{\phi_c'}{a^3}(\Phi' a + 2 a' \Phi) +\nonumber\\+ \frac{1}{a^2}(\square_{ij} - \Delta_{ij})\varphi \big]-
4\alpha \frac{\Phi}{a^2}(\phi_c')^2(\frac{\phi_c''}{a^2} - \gamma \frac{\phi_c' a'}{a^3} - \frac{a' \phi_c'}{a^3}) +
4\alpha \varphi' \frac{\phi_c'}{a^2}(\frac{\phi_c''}{a^2} - \gamma \frac{\phi_c' a'}{a^3} - \frac{a' \phi_c'}{a^3}) +
4\alpha \frac{\Phi(\phi_c')^2\phi_c''}{a^4} -\nonumber\\- 2\alpha \frac{\varphi' \phi_c'\phi_c''}{a^4} - 2\alpha \frac{\phi_c'\phi_c''\varphi'}{a^4} - 2\alpha\frac{(\phi_c')^2\varphi''}{a^4}+ 4\alpha  \frac{(\phi_c')^2\phi_c''\Phi}{a^4} - 4\alpha\frac{a'(\phi_c')^3\Phi}{a^5} + 2\alpha\frac{a'\varphi'(\phi_c')^2 }{a^5} + 2\alpha \frac{(\phi_c')^3}{a^2}(\frac{\Phi'}{a^2} - 2\frac{\Phi a'}{a^3})=\nonumber\\=
2\alpha\gamma\frac{(\phi_c')^2}{a^4}(\phi_c' \Phi'  + 4\Phi \frac{a'}{a}\phi_c' - 3\varphi' \frac{a'}{a}) - 2\alpha\frac{(\phi_c')^2}{a^4}\square_{ij}\varphi
\end{gather} 

\end{center}

And we must do the variation of the tensor of energy momentum:
\begin{center}
\begin{gather}
\delta {T}^{0}_{\ 0}= 2\alpha\frac{(\phi_c')^2}{a^2}\big[-\frac{2}{a^2}\phi_c''\Phi + \frac{\varphi''}{a^2} - \frac{\Delta_{ij}\varphi}{a^2} + \frac{\gamma}{a^2}(-\phi_c'\Psi' - 2\Phi\frac{a'}{a}\phi_c' - \frac{\varphi' a'}{a} ) -\frac{\varphi'a' }{a^3}  + \frac{4\Phi\phi_c' a'}{a^3} - \frac{\phi_c'}{a^3}(\Phi' a + 2 a' \Phi) +\nonumber\\+ \frac{1}{a^2}(\square_{ij} - \Delta_{ij})\varphi \big]-
4\alpha \frac{\Phi}{a^2}(\phi_c')^2(\frac{\phi_c''}{a^2} - \gamma \frac{\phi_c' a'}{a^3} - \frac{a' \phi_c'}{a^3}) +
4\alpha \varphi' \frac{\phi_c'}{a^2}(\frac{\phi_c''}{a^2} - \gamma \frac{\phi_c' a'}{a^3} - \frac{a' \phi_c'}{a^3}) +
4\alpha \frac{\Phi(\phi_c')^2\phi_c''}{a^4} -\nonumber\\ - 2\alpha \frac{\varphi' \phi_c'\phi_c''}{a^4} - 2\alpha\frac{(\phi_c')^2\varphi''}{a^4}-2\alpha\frac{\phi_c'\phi_c''\varphi'}{a^4} + 4\alpha  \frac{(\phi_c')^2\phi_c''\Phi}{a^4} - 4\alpha\frac{a'(\phi_c')^3\Phi}{a^5} + 2\alpha\frac{a'\varphi'(\phi_c')^2 }{a^5} + 2\alpha \frac{(\phi_c')^3}{a^2}(\frac{\Phi'}{a^2} - 2\frac{\Phi a'}{a^3})=\nonumber\\= 
2\alpha\gamma\frac{(\phi_c')^2}{a^4}(\phi_c' \Phi'  + 4\Phi \frac{a'}{a}\phi_c' - 3\varphi' \frac{a'}{a}) - 2\alpha\frac{(\phi_c')^2}{a^4}\square_{ij}\varphi\nonumber
\end{gather}
\begin{gather}
\delta T^{i}_{j} = \alpha g^{i}_{j}\frac{\phi_c'}{a^6} \big[ 4\phi_c''\varphi' a^2 + 2\varphi''\phi_c' a^2 - 6\phi_c' \varphi' a a' - 8\Phi \phi_c' a^2\phi_c'' + 8\Phi(\phi_c')^2 a' a  - 2(\phi_c')^2 a^2\Phi'\big]\nonumber
\end{gather}
\begin{gather}
\delta T^{0}_{\ i} = 2\alpha\ g^{00}\phi'_{c}\phi_{,i}\square\phi - \alpha\ g^{00}\phi'_{c}(\phi_{c,\rho}\phi_{,\sigma}g^{\sigma\rho})_{,i} - \alpha\varphi_{,i}g^{00}({\phi_{,\rho}}\phi_{,\sigma}g^{\sigma\rho})_{,0} = \nonumber\\ = \frac{2}{a^2}\alpha\phi'_{c}\varphi_{,i}(\frac{\phi_c''}{a^2} - \frac{\gamma}{a^2}\phi_c'  \frac{a'}{a} - \frac{a' \phi_c'}{a^3})- \alpha g^{00}(\phi_c' +\varphi')(\phi_{,\rho i}\phi_{,\sigma} g^{\sigma\rho} + \phi_{,\rho}\phi_{,\sigma i} g^{\sigma\rho} + \phi_{,\rho}\phi_{, \sigma} g^{\sigma\rho}_{\ \ ,i}) - \nonumber\\
- \alpha g^{00}\varphi_{,i}(\phi_{,\rho 0}\phi_{,\sigma} g^{\sigma\rho} + \phi_{,\rho}\phi_{,\sigma 0}g^{\sigma\rho} + \phi_{,\rho}\phi_{,\sigma} g^{\sigma\rho}_{\ \ ,0})= \nonumber\\ = \frac{2\alpha}{a^2}\phi_c'\varphi_{, i}(\frac{\phi_c''}{a^2} - \frac{\gamma}{a^2}\phi_c'\frac{a'}{a} - \frac{a' \phi_c'}{a^3}) - 2\alpha\frac{(\phi_c')^2}{a^4}\varphi_{,0i} + 2\alpha \frac{\Phi_{,i}}{a^4}(\phi_c')^3 + 2\alpha\frac{\varphi_{,i}}{a^2}\big[-\phi_{c,00}\frac{\phi_{c,0}}{a^2} + (\phi_{c,0})^2 \frac{a'}{a^3} \big], 
\end{gather}

where $g^{\sigma\rho}_{\ \  ,0}$ is a derivative of contravariant metric with respect to the conformal time.

\end{center}

So, the pertubed part of the Einstein equations read \cite{Kumar}:

\begin{gather}
\Delta\Phi - 3H(\Phi' + H\Phi) + 3K\Phi = \frac{\kappa}{2}a^2(\delta\epsilon_{dust} + \delta\epsilon_{rad}) 
+\frac{\kappa}{2}\big[-(\phi_c')^2\Phi + \phi_c' \varphi' + a^2\frac{dV}{d\phi}(\phi_c)\varphi+\nonumber\\
+2\alpha\gamma\frac{(\phi_c')^2}{a^2}(\phi_c'\Phi' - 3\varphi' \frac{a'}{a})
-2\alpha \frac{(\phi_c')^2}{a^2}\square_{ij}\varphi\big],\label{one}
\end{gather}

\begin{gather}\label{2}
\partial_i \Phi' + H\partial_i\Phi = \frac{\kappa}{2}\big[\phi_c'\partial_i\varphi + 2\alpha\phi_c' \partial_i\varphi(\frac{\phi_c''}{a^2} - \frac{\gamma}{a^2}\phi_c' \frac{a'}{a} - \frac{a'}{a}\phi_c')  - 2\alpha \frac{(\phi_c')^2}{a^2}\varphi_{,0i} + \frac{2\alpha \Phi_{,i}}{a^2}(\phi_c')^3 + 2\alpha \frac{\varphi_{,i}}{a^2}[-\phi_{c,00}\phi_{c,0} + (\phi_{c,0})^2\frac{a'}{a}\big],
\end{gather}

\begin{gather}
\frac{2}{a^2}\big\{\Phi'' + 3H\Phi' + \Phi(2\frac{a''}{a} - H^2 - K)\big\} 
=\kappa \big\{\delta p_{rad} - \frac{(\phi_c')^2\Phi}{a^2} + \frac{\phi_c' \varphi'}{a^2} - \frac{dV}{d\phi}\phi_c \varphi - \frac{\alpha\phi_c'}{a^6} [ 4\phi_c'' \varphi' a^2 + 2\varphi'' \phi_c' a^2 -  6\phi_c' \varphi' a a' -\nonumber\\- 8\Phi \phi_c' a^2\phi_c'' + 8\Phi (\phi_c')^2 a' a - 2(\phi_c')^2 a^2 \Phi']\big\}.\label{third}
\end{gather}

We obtain the field equation for the field $\phi$ by carrying the  variation of (\ref{action}) and the result reads: 

\begin{center}
\begin{gather}
-2\alpha(\square\phi)^2 + 2\alpha\nabla^{\mu}\nabla^{\nu}\phi\nabla_{\mu}\nabla_{\nu}\phi + 2\alpha \nabla^{\mu}\phi\nabla^{\nu}\phi R_{\mu\nu} -
\square\phi - \frac{dV}{d\phi}=0\label{EoM} 
\end{gather}

\end{center}

After plugging in the definition of the energy-momentum tensor, we have that 

\begin{gather}
\bar{\epsilon}_{\varphi} = -8\alpha\frac{(\phi_c')^3a'}{a^5} + \frac{2\alpha\phi_c''}{a^4}(\phi_c')^2 + \frac{1}{2a^2} (\phi_c')^2 + V(\phi_c),
\end{gather}

\begin{gather}
\bar{p}_{\varphi} = -6\alpha\frac{\phi_c''\phi_c'^2}{a^4} + 6 \alpha \frac{a'(\phi_c')^3}{a^5} + \frac{1}{2a^2} (\phi_c')^2 - V(\phi_c),
\end{gather}

where $\bar{\epsilon}_{\varphi}$ and $\bar{p}_{\varphi}$ are the energy density of the Galileon field at the background level.

Consequently, Friedmann and Raychaudhuri equations read: 
\begin{equation}\label{HH}
H^2 = \frac{\kappa a^2}{3} \big[ \bar{\epsilon}_{DUST} + \bar{\epsilon}_{RAD} + \frac{1}{2a^2} (\phi_c')^2 + V(\phi_c) - 8\frac{\alpha a^{'}(\phi_c')^3}{a^5} + 2\alpha(\phi_c')^2\frac{\phi_c''}{a^4}\big] - K
\end{equation}

\begin{equation}\label{HPRIME}
H' = -\frac{\kappa a^2}{6}\big[\bar{\epsilon}_{DUST} + 2\bar{\epsilon}_{RAD} - 16\alpha \frac{(\phi_c')^2\phi_c''}{a^4} + 10\alpha \frac{a^{'}}{a^5}(\phi_c')^3 + 2 \frac{(\phi_c')^2}{a^2} - 2V(\phi_c)\big]
\end{equation}

We continue now in two steps: we compute the derivative of $\Phi' + H\Phi$ and we plug the second Einstein equation (\ref{2}) to the third Einstein equation (\ref{third}):

\begin{equation}
\Phi' + H\Phi = \frac{\kappa}{2}\big\{\phi_c'\varphi + 2\alpha \phi_c'\varphi(-\frac{\gamma\phi_c' a'}{a^3})-2\frac{\alpha(\phi_c')^2a'\varphi}{a^3}-2\alpha\frac{(\phi_c')^2}{a^2}\varphi' + 2\alpha\frac{\Phi}{a^2}(\phi_c')^3 + 2\alpha \frac{\varphi}{a^2}[(\phi_c')^2\frac{a'}{a}]\big\}
\end{equation}

\begin{gather}
\Phi'' + H' \Phi + H\Phi' = \frac{\kappa}{2a^4}\{\phi_c'' a^4\varphi + \phi_c' a^4\varphi' -12\alpha\phi_c''\phi_c' a a' \varphi - 2\alpha(\phi_c')^2\varphi' a' a -6\alpha(\phi_c')^2a'' a\varphi + \nonumber\\ + 18\alpha(\phi_c')^2(a')^2\varphi - 2\alpha(\phi_c')^2\varphi'' a^2 - 4\alpha \phi_c'\phi_c''\varphi' a^2+ 2\alpha\Phi'(\phi_c')^3a^2 + 6\alpha\Phi(\phi_c')^2\phi_c'' a^2 - 4\alpha\Phi(\phi_c')^3a'a\}
\end{gather}

So after plugging to the third Einstein equation (\ref{third}):
\begin{gather}
\frac{2}{a^2}\big\{\frac{\Phi a^2}{3}(\frac{-3}{2}\bar{\epsilon}_{DUST} - 2\bar{\epsilon}_{RAD} + 6\alpha\frac{(\phi_c')^2\phi_c''}{a^4} + 3\alpha \frac{a'(\phi_c')^3}{a^5} - \frac{3}{2}\frac{(\phi_c')^2}{a^2}) + \nonumber\\ + 
H[\phi_c'\varphi -6\alpha\phi_c'\varphi\frac{\phi_c'a'}{a^3} - 2\alpha\frac{(\phi_c')^2 \varphi'}{a^2} + 2\alpha\Phi\frac{(\phi_c')^3}{a^2}]+ 
\frac{1}{2}\big[\frac{1}{a^4}(\phi_c''a^4\varphi + \phi_c' \varphi'a^4 - 12\alpha(\phi_c'')a \phi_c' a'\varphi - \nonumber\\ - 2\alpha (\phi_c')^2\varphi' a' a - 4\alpha\phi_c'\phi_c'' a^2\varphi' - 6\alpha(\phi_c')^2a''a\varphi + 18\alpha(\phi_c')^2(a')^2\varphi - 2\alpha (\phi_c')^2\varphi'' a^2 + 2\alpha (\phi_c')^3a^2 + \nonumber\\ \big[-H\Phi + \frac{\kappa}{2}\big(\phi_c'\varphi-6\alpha\phi_c'\varphi(\frac{\phi_c'a'}{a^3} ) - 2\alpha \frac{(\phi_c')^2}{a^2}\varphi' + 2\alpha\Phi\frac{(\phi_c')^3}{a^2})+  6 \alpha\Phi(\phi_c')^2\phi_c'' a^2 - 4\alpha \Phi(\phi_c')^3a' a\big]\big\} + (\phi_c')^2\frac{\Phi}{a^2} + \frac{dV}{d\phi}\varphi - \frac{\phi_c'\varphi'}{a^2}+ \nonumber\\+ \alpha\frac{\phi_c'}{a^6}(4\phi_c''\varphi' a^2 + 2\varphi''\phi_c'a^2 - 6\phi_c'\varphi' a a'  - 8\Phi\phi_c'\phi_c'' a^2 + 8\Phi(\phi_c')^2a'a -\nonumber\\- 2(\phi_c')^2a^2\big\{-H\Phi +\frac{\kappa}{2}\big[\phi_c'\varphi + 2\alpha\phi_c'\varphi\big(-3\frac{\phi_c' a'}{a^3 }- 2\frac{\alpha(\phi_c')^2}{a^2}\varphi' + 2\alpha\Phi\frac{(\phi_c')^3}{a^2}\big)\big]\big\} = \frac{\delta\epsilon_{rad}}{3}
 \end{gather}

We further simplify this equation as: 

\begin{gather}
\frac{\delta\epsilon_{rad}}{3} = -\frac{1}{3a^6}[(-6a'\phi_c'a^3 - 3\frac{dV}{d\phi}a^6 - 3\phi_c'' a^4 + 12\alpha\phi_c'' \phi_c' a a' + 18 \alpha (\phi_c')^2 a'' a - 18\alpha(\phi_c')^2(a')^2\varphi - 30\alpha\Phi(\phi_c')^3 a' a -\nonumber\\- 6 \alpha\Phi (\phi_c')^2\phi_c'' a^2 + (36\alpha(\phi_c')^2 a' a)\varphi'  + 3 \Phi a^6\bar{\epsilon}_{dust} + 4\Phi a^6\bar{\epsilon}_{rad}]
\end{gather}

So, after inserting the equation of motion, we get the following equation: 

\begin{gather}
\frac{\delta\epsilon_{rad}}{3} = \frac{-1}{3a^6}[(36\alpha H\phi_c'\phi_c'' + 18\alpha H^2(\phi_c')^2a^2 + 18\alpha H'(\phi_c')^2a^2+
+12\alpha\phi_c''a\phi_c'a' + 18\alpha(\phi_c')^2a''a- 
18\alpha(\phi_c')^2(a')^2)\varphi -\nonumber\\- 30\alpha\Phi(\phi_c')^3a' a- 6\alpha\Phi(\phi_c')^2\phi_c''a^2 + 
36\alpha(\phi_c')^2a' a\varphi' + 3\Phi a^6 \bar{\epsilon}_{dust} + 4\Phi a^6\bar{\epsilon}_{rad}]\nonumber
\end{gather}

We can plug this expression into the first Einstein equation (\ref{one}): 

\begin{gather}
\frac{2}{\kappa}\Delta\Phi + \frac{6K}{\kappa}\Phi - a^2(\frac{\delta\rho c^2}{a^3} + 3\bar{\rho} \frac{\Phi}{a^3}) - 36\alpha H \frac{(\phi_c')^3\Phi}{a^2} - \frac{6\alpha}{a^2}\Phi (\phi_c')^2\phi_c'' + 6\frac{\kappa\alpha^2(\phi_c')^6\Phi}{a^4} + (\phi_c')^2\Phi +3\Phi a^2 \bar{\epsilon}_{dust} + 4\Phi a^2\bar{\epsilon}_{rad} -\nonumber\\- \big[3H\phi_c'  + 72 \frac{\alpha}{a^2} H\phi_c' \phi_c'' + 36\alpha (\phi_c')^2\frac{a''}{a^3} + a^2 \frac{dV}{d\phi}\big]\varphi + \big[\frac{3\alpha(\phi_c')^4}{a^2} - 18\frac{\alpha^2H (\phi_c')^5\kappa\varphi}{a^4}\big] + (\frac{60\alpha}{a^2} H (\phi_c')^2 - \phi_c')\varphi' -\nonumber\\- \frac{6\alpha^2(\phi_c')^5}{a^4}\kappa\varphi' + 2\alpha \frac{(\phi_c')^2\square_{ij}\varphi}{a^2}=0
\end{gather} 

\section{Results}
Similarly to \cite{Kumar} we suppose that $\Omega = \Omega(r)$ (where $\Phi = \Omega/r$ and $\Omega$ is a function of $a$ and spatial coordinates), which means that $\phi_c' = const.$: 

$$
\phi_c(\eta) = \beta\eta + \omega,
$$
where $\omega$ and $\beta$ are constants. Then we get from EoM (\ref{EoM}) that 
\begin{equation}\label{EoM}
6\alpha\beta^2a'' + 2 a^2a'\beta + \frac{dV}{d\phi}a^5 = 0.\nonumber
\end{equation}

Now

\begin{gather}\label{VEQ}
V(\eta) = \frac{\beta^2}{a^2} + V_{\infty} + \alpha f(a)
\end{gather}

and the goal is to obtain the dependence $f(\eta)$ in this previous relation, when we know the dependence $V(a) = \frac{\beta^2}{a^2} + V_{\infty}$ for the pure scalar field, \cite{Kumar}. So we take the equation

\begin{gather}
6 \alpha\beta^3\frac{a''}{a^3} + 2\frac{a'}{a}\beta^2 + a^2 V' = 0
\end{gather}

and we take the derivate of the previous equation (\ref{VEQ}) with respect to $\eta$ and obtain:

\begin{gather}
V' a^2 = -2\beta^2 \frac{a'}{a} + \alpha \frac{df}{da}a' a^2
\end{gather}

Now we combine these two equations: 

\begin{gather}
6\alpha\beta^3\frac{a''}{a} + \alpha  \frac{df}{da}a' a^4 = 0
\end{gather}

We use that $\frac{a''}{a} = H' + H^2$ and substitute $H$ and $H'$ from (\ref{HH}) and (\ref{HPRIME}). We finally obtain that: 
\begin{gather}\label{alpha}
2\alpha\beta^3\kappa[\frac{\bar{\rho}c^2}{2a^3} + 3 \frac{\beta^2}{2a^2} + 2V_{\infty} + 2\alpha f(a) - 13\alpha\frac{a'\beta^3}{a^5} ] + \alpha \frac{df}{da}a'  a^2 = 0
\end{gather}

We should solve this equation for $f(a)$ and we use the Taylor expansion. We must find the solution of the equation (\ref{HH}) for $H$: 

\begin{gather}
 0 = H^2 + \frac{8\kappa\alpha\beta^3}{3a^2}H - \frac{\kappa a^2}{3}[\bar{\epsilon}_{dust} + \bar{\epsilon}_{rad} + \frac{3\beta^2}{2a^2} + \alpha f(a) +  V_{\infty}]
\end{gather}

This is a quadratic equation. We ignore the negative solution like unphysical  and we obtain for the positive one the following expression: 

\begin{gather}
H(\alpha) = -\frac{4\alpha\beta^3\kappa}{3a^2} + \sqrt{\frac{16}{9}(\frac{\alpha\beta^3\kappa}{a^2})^2 + \kappa \frac{a^2}{3}(\bar{\epsilon}_{DUST} + \bar{\epsilon}_{RAD} + \frac{3\beta^2}{2a^2} + \alpha f(a) + V_{\infty})} 
\end{gather} 

We expand this expression in $\alpha$ and we get: 

\begin{gather}
H(\alpha)|_{\alpha = 0} = \sqrt{\frac{\kappa a^2}{3}(\bar{\epsilon}_{DUST} + \bar{\epsilon}_{RAD} + \frac{3\beta^2}{a^2} + V_{\infty}) }+ [\frac{-4\kappa\beta^3}{3a^2} + \frac{\kappa a^2 f(a)}{6\sqrt{\frac{\kappa a^2}{3}(\bar{\epsilon}_{DUST} + \bar{\epsilon}_{RAD} + \frac{3\beta^2}{2a^2} + V_{\infty}}}] + O(\alpha^2)
\end{gather}

We obtain for the equation (\ref{alpha}) in the first order approximation 

\begin{gather}
0  = 2\alpha\beta^3\kappa[\frac{\bar{\rho}c^2}{2a^3} + \frac{3\beta^2}{2a^2} + 2 V_{\infty}] + \alpha \frac{df(a)}{da}\sqrt{\frac{\kappa a^2}{3}(\bar{\epsilon}_{dust} + \bar{\epsilon}_{rad} + \frac{3\beta^2}{2a^2} + V_{\infty}) a^3}
\end{gather}
We see after some simplifications that 

\begin{equation}
\frac{df(a)}{da} = \frac{-2\beta^3\kappa[\bar{\rho} c^2{2} + \frac{3\beta^2}{2}a +2 V_{\infty} a^3]} {a^5\sqrt{H_0^2\Omega_r + \frac{\kappa\bar{\rho}c^2}{3}a + \frac{\kappa\beta^2}{2}a^2 + \frac{V_{\infty}\kappa a^4}{3}}}
\end{equation}

So $f(a)$ behaves like matter: 

\begin{gather}\label{a}
f(a)\sim \frac{1}{a^3}
\end{gather}

\section{Conclusion}

We were studying Galileon cosmologies in this work. We used the mechanical approach, because we worked deep inside the cell of uniformity 150-370 Mpc and in the late universe. We filled the Universe with minimally coupled Galileon field and also dust like matter in the form of discrete distributed galaxies and groups of galaxies. We further included radiation. 

All types of inhomogeneities have non-relativistic velocities. Different types of inhomogeneities do not not run away massively during the universe evolution in this case. Fluctuations of the energy density of such perfect fluids are usually concentrated around the inhomogeneities of dust like matter. Therefore we call those perfect fluids coupled \cite{Perfect}. They can even screen the gravitational potential of galaxies \cite{Alvina} or they can play a role of dark matter flattening the rotation curves of dwarf galaxies \cite{Brilenkov}.  In the present work, we have investigated the possibility for Galileon scalar field to be coupled with Galaxies in the late universe. For such Galileon scalar fields to exist, we have shown that they have to meet certain conditions. At the background level, such scalar field behaves as a three component perfect fluid: a network, which mimics a behavior of frustrated cosmic strings with the EoS parameter $w=-1/3$, a cosmological constant and some matter component, (\ref{a}). This is the main result of our work.

\section{Acknowledgement}

I would like to thank Mariam Bouhmadi-L\'opez for discussions on the topic and for the possibility to stay at the university UPV/EHU in Bilbao in 2017 and 2018.


\begin{thebibliography}{9}
\bibitem{Weinberg}
  S.~Weinberg, The cosmological constant problem, Rev.Mod.Phys. 61, 1 , 1989, arXiv: 0005265 [astro-ph]
  
\bibitem{Starobinsky} 
V.~Sahni, A.~Starobinsky, The case for a positive cosmological Lambda-term, Int.J.Mod.Phys. D9: 373-444, 2000, arXiv: 9904398 [astro-ph] 
\bibitem{Amendola} L.~Amendola, S.~Tsujikawa, Dark energy, Cambridge University Press, 2010
\bibitem{Lopez} M.~Bouhmadi-L\'opez, K.~Sravan Kumar, J.~Marto, J.~Morais and A.~Zhuk, K-essence model from the mechanical approach point of view: coupled scalar field and the late cosmic acceleration, JCAP 07, 050, 2016, arXiv: 1605.03212v2  [gr-qc]

\bibitem{Faraoni} S.~Capozziello, V.~Faraoni, Beyond Einstein Gravity: A Survey of Gravitational Theories for Cosmology and Astrophysics, Fundam.Theor.Phys.170, 2010
\bibitem{Capozziello} S.~Capozziello, M.~De Laurentis, Extended Theories of Gravity, Physics Reports 509 (4-5): 167-321, 2011, arXiv: 1108.6266 [gr-qc]  
\bibitem{Chimento} M.~Bouhmadi-López, L.~P.~Chimento, Phys. Rev. D 82, 103506, arXiv:1007.4141 [astro-ph.CO]

\bibitem{Rattazzi} A.~Nicolis, R.~Rattazzi, E.~Trincherini, The galileon as a local modification of gravity, Phys.Rev. D 79, 064036, 2009, 
arXiv:0811.2197 [hep-th]
\bibitem{Felice} A.~De Felice, S.~Tsujikawa, Generalized Galileon cosmology, Phys. Rev. D84: 124029, 2011, arXiv: 1008.4236 [hep-th]
\bibitem{Khoury}J.~Khoury, A.~Weltman, Chameleon cosmology, Phys.Rev. D 69, 044026, 2004,  arXiv: 0309411 [astro-ph] 
\bibitem{Babichev} E.~Babichev, C.~Deffayet, An introduction to the Vainshtein mechanism, Class. Quantum Grav. 30, 184001, 2013, 
arXiv:1304.7240 [gr-qc]

\bibitem{Miguel} J.~M.~Ezquiaga, M.~Zumalac\'arregui, Dark energy after GW170817: dead ends and the road ahead, Phys. Rev. Lett. 119, 251304, 2017,  arXiv:1710.05901v2 [astro-ph.CO]  
\bibitem{Mukhanov} V.~F. Mukhanov, H.~A.~Feldman, R.~H.~Brandenberger, Theory of cosmological perturbations, Physics Reports 215, 203, 1992 
\bibitem{Eingorn} M.~Eingorn, First-order cosmological perturbations engendered by point-like masses, Astroph. J. 825, no.2, 84, arXiv: 1509.03835 [gr-qc] 
\bibitem{Zhuk} M.~Eingorn, A.~Zhuk, Remarks on mechanical approach to observable Universe, JCAP 05, 024, 2014, arXiv: 1309.4924 [astro-ph.CO]
\bibitem{Alexander} M.~Eingorn, A.~Zhuk, Hubble flows and gravitational potentials in observable Universe, JCAP 09, 026, 2012, arXiv: 1205.2384  [astro-ph.CO] 
\bibitem{Burgazli} A.~Burgazli, M.~Eingorn, A.~Zhuk, Rigorous theoretical constraint on constant negative EoS parameter w and its effect for the late Universe, Eur. Phys. J. C 75, 118, 2015
\bibitem{Brilenkov} M.~Brilenkov, M.~Eingorn, L.~Jenkovszky, A.~Zhuk, Scalar perturbations in cosmological models quark nuggets, Eur. Phys. J. C 74, 3011, 2014, arXiv: 1310.4540 [astro-ph.CO] 
\bibitem{Akarsu} \"O.~Akarsu, M.~Bouhmadi-L\'opez, M.~Brilenkov, R.~Brilenkov, M.~Eingorn, A.~Zhuk, Are dark energy models with variable EoS parameter w compatible with the late inhomogeneous Universe?, JCAP 07, 038, 2015, arXiv: 1502.04693 [gr-qc] 
\bibitem{Morais} M.~Bouhmadi-L\'opez, M.~Brilenkov, R.~Brilenkov, J.~Morais, A.~Zhuk, Scalar perturbations in the late Universe: viability of the Chaplyagin gas model, JCAP 12, 037, 2015 [gr-qc]
\bibitem{Perfect} A.~Zhuk, Perfect fluids coupled to inhomogeneities in the late Universe, Gravitation and Cosmology, 22, 159, 2016, arXiv: 1601.01939 [gr-qc]  
\bibitem{Kumar} A.~Burgazli, A.~Zhuk, J.~Morais, M.~Bouhmadi-L\'opez, K.~Sravan Kumar: Coupled scalar fields in the late Universe, The mechanical approach and the late cosmic acceleration, JCAP 09, 045, 2016, arXiv: 1512:03819 [gr-qc]
\bibitem{Albarran} I.~Albarran, M.~Bouhmadi-L\'opez, J.~Morais, Cosmological perturbations in an effective and genuinely phantom dark energy Universe, Phys. Dark Univ. 16, 94-108, 2017, arXiv: 1611.00392  [astro-ph.CO] 
\bibitem{Quiros} I.~Quiros, R.~Garc\'ia-Salcedo, T.~Gonzalez, F.~Antonio Horta-Rangel, J.~Saavedra, Brans-Dicke Galileon and the variational principle,  arXiv:1605.00326v2 [gr-qc]   
\bibitem{Alexandra} M.~Eingorn, A.~Kudinova, A.~Zhuk, Dynamics of astrophysical objects against the cosmological background, JCAP 04, 010, 2013, arXiv:1211.4045 [astro-ph.CO]
\bibitem{JNa} M.~ Eingorn, J.~Nov\'ak, A.~Zhuk, arXiv: 1401.5410, f(R) gravity: scalar perturbations in the late Universe, The European physical journal C 74, 3005, 2014, 
arXiv:1401.5410 [astro-ph.CO]
\bibitem{Alvina} A.~Burgazli, M.~Eingorn and A.~Zhuk, Rigorous theoretical constraint on constant negative EoS parameter $w$ and its effect for the late Universe, Eur. Phys. J. C 75 (2015) 118, arXiv:1301.0418  [astro-ph.CO]
\bibitem{Brilenkov} M.~Brilenkov, M.~Eingorn, L.~Jenkovszky and A.~Zhuk, Scalar perturbations in cosmological models with quark nuggets, Eur. Phys. J. C 74 (2014) 3011 arXiv:1310.4540 [astro-ph.CO]
\end{thebibliography}
\end{document}